\documentclass[10pt,a4paper,twoside]{article}

\usepackage{indentfirst}            % Required!
\usepackage{graphicx}               % Required if figures
\usepackage{amsgen,amsfonts,amssymb,amsbsy} % Enlarge math symbols set

\newenvironment{resum}{\begin{quote}\small}{\end{quote}}

\newcommand{\bfsf}[1]{\textsf{\textbf{#1}}}
\newcommand{\lf}[2]{\mbox{\Large $\frac{#1}{#2}$}}

\begin{document}

\thispagestyle{plain}       % Remove headings in first page

\begin{center}

%  Title; use linebreaks with "\\" if necessary:

{\LARGE\bfsf{Black Hole Entropy: Linear Or Quadratic?}}

\bigskip

%  Author list:

\textbf{Antonio Alfonso-Faus }

%  Affiliation. State only name and Institution of authors:

\emph{E.U.I.T. Aeron\'autica,  Spain}\/

\end{center}

\medskip

%  Place abstract here:

\begin{resum}
The entropy $S$ of any mass $M$ can be interpreted as a linear relation with mass, $S \simeq
kM/m_g$ an extensive property with $m_g$ the mass of the quantum of gravity. One can extend
this relation to black holes and then we get a different result, as compared with the
``standard'' relation $S \simeq k/\hbar c \; GM^2$ which is quadratic in the mass. We discuss
both approaches and apply it to cosmology.

\end{resum}

\bigskip

%  ********************************************************
%       Contribution proper begins here:
%  ********************************************************

\section{Introduction}

We will refer to three concepts already present in the scientific literature: the gravity
quanta [1], the gravitational cross section  and the gravitational entropy [2].  The first is
a step in the direction of introducing quantum mechanical concepts into general relativity and
cosmological treatments. From general relativity we know that the gravitational field (energy)
must be unlocalized, and this must be the case for the gravity quanta. We impose then the
geometrical condition that the Compton wavelength of the gravity quanta of mass $m_g$ must be
of the order of the size of the seeable Universe $ct$, where $t$ is the age of the Universe:
\begin{equation}
\lf{\hbar}{m_g \,c} \simeq ct \qquad \Longrightarrow \qquad m_g \simeq \lf{\hbar}{c^2 t} \simeq %
2\cdot 10^{-66} \;\mbox{gr}
\end{equation}

The second concept, the gravitational cross section, is derived from Newton's gravitational
force expressed as an interchange of momenta between gravitational masses: $M$ gravitating on
$m$. The mass $M$ has acquired through the age of the Universe $t$ a ``well'' of momentum
$Mc$. This can act on an effective gravitational cross section $\sigma_g$ of the mass $m$ in
accordance with the geometrical proportion $\sigma_g/4\pi r^2$. This mechanical definition
gives
\begin{equation}
  \lf{GMm}{r^2}\simeq \lf{Mc}{t}\cdot \lf{\sigma_g}{4\pi r^2}
\end{equation}
and then
\begin{equation}
\sigma_g \simeq 4\pi \lf{Gm}{c^2}\cdot ct
\end{equation}
We see that the gravitational cross section of a mass m is of the order of the product of its
gravitational radius times the size of the seeable Universe. As a reference, particularizing
for a proton, its gravitational cross section is of the order of its area, as defined by the
square of its Compton wavelength. On the other hand, the gravitational cross section for the
whole Universe of mass $M_u$ is the square of its size $ct \simeq GM_u/c^2$ as given by Mach's
principle in the way we will interpret later. Then one has [2]
\begin{equation}
  \sigma_g \simeq 4\pi \lf{Gm}{c^2}\cdot \lf{GM_u}{c^2}
\end{equation}
The conclusion is that the gravitational cross section of any mass $m$ is of the order of the
product of two gravitational radii as given by its mass and the mass of the Universe.

Following an approach similar to Hawking [3] and Bekenstein [4], and using our gravitational
cross section concept (4) we define entropy $S$ of any mass $m$ as
\begin{equation}
  \lf{S}{k}\simeq \lf{\sigma_g}{A}
\end{equation}
where $A$ is the Planck's area. Using (4) we get then
\begin{equation}
  \lf{S}{k}\simeq 4\pi \lf{Gm}{c^2}\cdot ct \cdot \lf{c^3}{G\hbar} = %
  4\pi \lf{mc^2 t}{\hbar} = 4\pi \lf{m}{m_g}
\end{equation}

The dimensionless entropy $S/k$ of a mass $m$ is the number of gravity quanta emitted: this
definition of entropy is an extensive property and is linear with mass. This is a
gravitational entropy.

\section{Standard Black Hole Entropy Versus the New Approach}

The Hawking [3] and Bekenstein [4] black hole entropy is given by
\begin{equation}
 \lf{S}{k}\simeq \left( \lf{Gm}{c^2}\right)^2 \cdot \lf{c^3}{G\hbar} = %
 \lf{1}{\hbar c} Gm^2
\end{equation}
which is quadratic in the mass and it has no time dependence. If we apply it to the mass of
the Universe, considering it as a kind of black hole, we get
\begin{equation}
 \lf{S}{k}\simeq \lf{1}{\hbar c} GM_u^2
\end{equation}
Our entropy definition (6) implies proportionality with mass and cosmological time. Our black
hole entropy is from (6)
\begin{equation}
 \lf{S}{k}\simeq 4\pi \lf{mc^2 t}{\hbar}
\end{equation}
and for the Universe
\begin{equation}
  \lf{S}{k}\simeq 4\pi \lf{M_u c^2 t}{\hbar}
\end{equation}
again linear with mass and cosmological time. We will transform this relation taking into
account a definition of Mach's principle to include cosmology.

\section{Mach's Principle}

We interpret Mach's Principle by saying that the gravitational potential energy of any mass
$m$, with respect to the rest of the Universe, is its rest energy (which can be related to one
form of the Equivalence Principle). Hence
\begin{equation}
  \lf{GM_u m}{ct} \simeq mc^2
\end{equation}
i.e.
\begin{equation}
 \lf{GM_u}{c^2}\simeq ct
\end{equation}

Then  using this we get our black hole gravitational entropy from (10)
\begin{equation}
 \lf{S}{k}\simeq \lf{4\pi}{\hbar c}\,GM_u^2
\end{equation}
which is the same as the Hawking-Bekenstein relation (8) for the Universe.

\section{Conclusions}

The gravitational entropy of a mass m can be defined as the number of gravity quanta emitted.
It is proportional to mass and cosmological time and therefore it is an extensive property,
something desired for the concept of entropy.

This entropy can be applied to BLACK HOLES, and the result is in contrast with the
Hawking-Bekenstein relation (that is quadratic in mass).

For the cosmological case, for the Universe using Mach's Principle, both approaches give the
same result.

The gravitational entropy of the Universe, as we have defined, is then
\begin{equation}
 \lf{S}{k}\simeq \lf{GM_u^2}{\hbar c} \simeq 10^{122}
\end{equation}
and we conclude that this number is so huge because the Universe is very old.

%\section{References}

\end{document}